\documentclass[aps,prl,twocolumn,showpacs,10pt,superscriptaddress,preprintnumbers,nofootinbib]{revtex4-1}
\usepackage{epsfig,amssymb,amsmath,psfrag,epstopdf,color}
\pdfoutput=1
\usepackage{soul}
\usepackage{graphicx}
\usepackage{paralist}
\usepackage{hyperref}
\allowdisplaybreaks
\def\beq{\begin{equation}}
\def\eeq{\end{equation}}
\def\bsp#1\esp{\begin{split}#1\end{split}}
\newcommand{\be}{\begin{equation}}
\newcommand{\ee}{\end{equation}}
\newcommand{\bea}{\begin{eqnarray}}
\newcommand{\eea}{\end{eqnarray}}

\def\Fig#1{Fig.~{\ref{#1}}}

\def\ksl{\not{\hbox{\kern-2.3pt $k$}}}

\def\spa#1.#2{\left\langle#1\,#2\right\rangle}
\def\spb#1.#2{\left[#1\,#2\right]}
\def\lor#1.#2{\left(#1\,#2\right)}
\def\sand#1.#2.#3{%
\left\langle\smash{#1}{\vphantom1}^{-}\right|{#2}%
\left|\smash{#3}{\vphantom1}^{-}\right\rangle}

\usepackage{xcolor}

\begin{document}

\preprint{MIT--CTP 5513}
\preprint{LA-UR-23-22542}
\title{Imaging Cold Nuclear Matter with Energy Correlators}

\author{Kyle Devereaux}
\email{devereauxk@berkeley.edu}
\affiliation{Nuclear Science Division, Lawrence Berkeley National Laboratory, Berkeley, CA 94720}

\author{Wenqing Fan}
\email{wenqing@lbl.gov}
\affiliation{Nuclear Science Division, Lawrence Berkeley National Laboratory, Berkeley, CA 94720}

\author{Weiyao Ke}
\email{weiyaoke@gmail.com}
\affiliation{Theoretical Division, Los Alamos National Laboratory, Los Alamos, NM 87545}

\author{Kyle Lee}
\email{kylel@mit.edu}
\affiliation{Center for Theoretical Physics, Massachusetts Institute of Technology, Cambridge, MA 02139}

\author{Ian Moult}
\email{ian.moult@yale.edu}
\affiliation{Department of Physics, Yale University, New Haven, CT 06511}

\begin{abstract}
The future electron-ion collider (EIC) will produce the first-ever high energy collisions between electrons and a wide range of nuclei, opening a new era in the study of cold nuclear matter. 
Quarks and gluons produced in these collisions will propagate through the dense nuclear matter of nuclei, imprinting its structure into subtle correlations in the energy flux of final state hadrons.
In this \emph{Letter}, we apply recent developments from the field of jet substructure, namely the energy correlator observables, to decode these correlations and provide a new window into nuclear structure.
The energy correlators provide a calibrated probe of the scale dependence of vacuum QCD dynamics, enabling medium modifications to be cleanly imaged and interpreted as a function of scale.
Using the eHIJING parton shower to simulate electron-nucleus collisions, we demonstrate that the size of the nucleus is cleanly imprinted as an angular scale in the correlators, with a magnitude that is visible for realistic EIC kinematics.
Remarkably, we can even observe the size difference between the proposed EIC nuclear targets ${}^3$He, ${}^4$He, ${}^{12}$C, ${}^{40}$Ca, ${}^{64}$Cu, ${}^{197}$Au, and ${}^{238}$U, showing that the energy correlators can image femtometer length scales using asymptotic energy flux.
Our approach offers a unified view of jet substructure across collider experiments, and provides numerous new theoretical tools to unravel the complex dynamics of QCD in extreme environments, both hot and cold.
\end{abstract}

\maketitle

\emph{Introduction.}---The future electron-ion collider (EIC) will provide the first electron-nucleus collisions at $\sqrt{s}$ up to 90 GeV, for a wide variety of nuclei \cite{osti_1765663,Accardi:2012qut,AbdulKhalek:2021gbh}. This presents a unique opportunity to study a broad range of phenomena in cold nuclear matter, ranging from parton energy loss and transport phenomena, to in-medium transverse-momentum broadening, and medium-modified hadronization. Additionally, the clean nature of the electron probe makes electron-ion collisions a simple system with a static nuclear medium that can serve as a foundation for understanding the more complex case of heavy-ion collisions \cite{Connors:2017ptx,Busza:2018rrf,Cunqueiro:2021wls,Apolinario:2022vzg}.  

As with any collider experiment, the key to success is extracting the details of the interactions of quarks and gluons with the nuclear matter 
from asymptotic observables measured on hadrons.
This is complicated by the complexity of the perturbative interactions of quarks and gluons, and the hadronization process in Quantum Chromodynamics (QCD). However, this complexity also represents an opportunity: energetic sprays of final state hadrons take the form of emergent multi-scale objects called jets, allowing intrinsic scales of the nuclear medium to be imprinted into scales of the jet. The study of the detailed internal structure of jets as a means to understand the underlying microscopic collision is referred to as jet substructure \cite{Larkoski:2017jix,Kogler:2018hem}, and its importance for the success of the EIC program has been emphasized in a number of recent studies \cite{Arratia:2019vju,Page:2019gbf,Li:2019dre,Li:2020zbk,Li:2020rqj,Li:2021gjw,Abelof:2016pby,Boughezal:2018azh,Gehrmann:2018odt,Dumitru:2018kuw,Liu:2018trl,Aschenauer:2019uex,Kang:2020fka,Li:2020bub,Arratia:2020azl, Kolbe:2020tlq,Ru:2023ars}.

\begin{figure}
\includegraphics[scale=0.24]{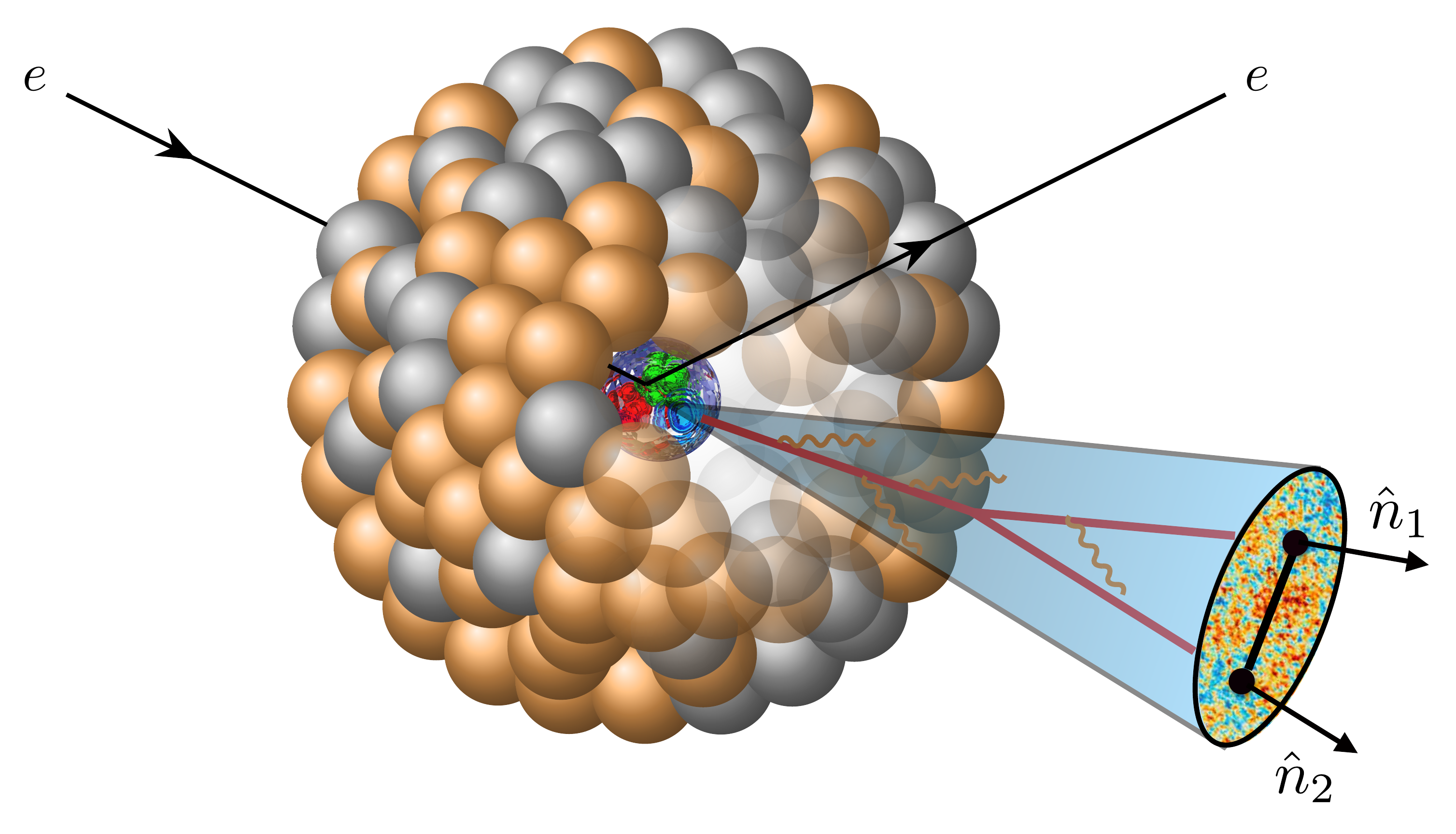}
\caption{A parton knocked out of a nucleon propagates a distance $\tau \sim 1/p_T \theta^2$, thereby directly imprinting nuclear time scales into angular scales of the two-point correlator in jet substructure.}
\label{fig:intro}
\end{figure}

Numerous spectacular recent advances in the understanding and analysis of jets, ranging from new machine learning (ML) approaches to unfold hyperdimensional data \cite{Andreassen:2019cjw}, to techniques enabling calculations on tracks \cite{Chen:2022pdu,Chen:2022muj,Jaarsma:2022kdd,Li:2021zcf}, to new approaches for performing perturbative calculations \cite{Kologlu:2019mfz,Chen:2021gdk,Chen:2022jhb,Chang:2022ryc}, allow one to re-imagine the future of jet substructure at the EIC. Central to recent developments in jet substructure has been the use of energy correlators \cite{Basham:1979gh,Basham:1978zq,Basham:1978bw,Basham:1977iq,Chen:2020vvp}, which measure statistical correlations in the energy flux within a jet, see \Fig{fig:intro}. In addition to their theoretical properties, these observables allow the formation of jets to be imaged as a function of scale, making them ideal for nuclear physics applications. This feature of the energy correlators has been illustrated for imaging the hadronization transition \cite{Komiske:2022enw}, measuring the top quark mass \cite{Holguin:2022epo}, observing intrinsic mass scales of heavy quarks before hadronization \cite{Craft:2022kdo}, resolving the scales of the quark-gluon plasma \cite{Andres:2022ovj,Andres:2023xwr}, and identifying the saturation scale in the color glass condensate~\cite{Liu:2022wop,Liu:2023aqb,Cao:2023rga}.

\begin{figure}
\includegraphics[scale=0.22]{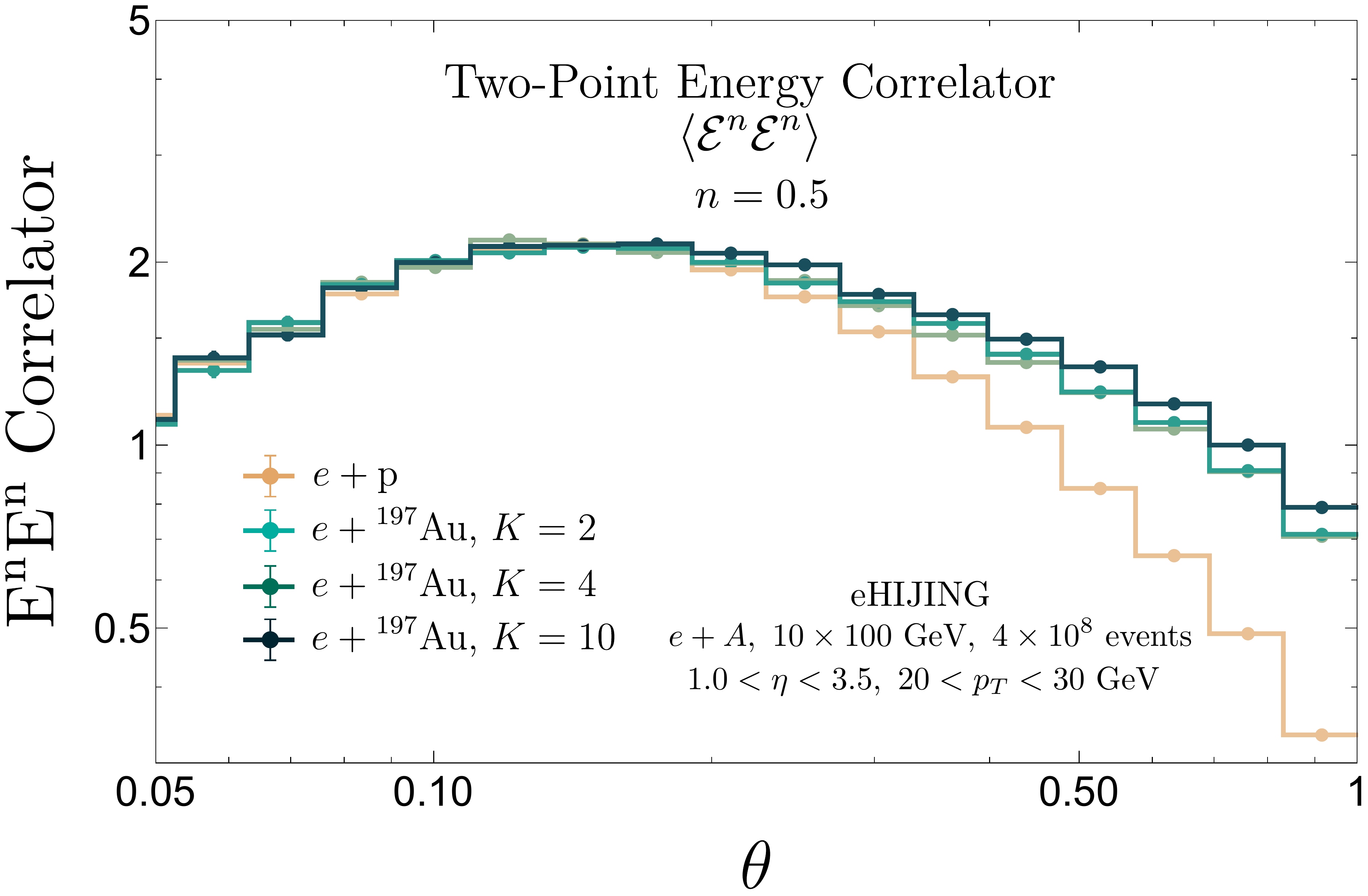}
\caption{The EEC, $\langle\mathcal{E}^n \mathcal{E}^n\rangle$, for $n=0.5$, and different values of the jet transport parameter, parameterized by the variable $K$ described in the text. The curves are normalized in the region $\theta\ll \theta_L$, where no medium modification is expected. A clear modification is seen at large angles.
}
\label{fig:raw_distros}
\end{figure} 

In this \emph{Letter}, we initiate a study of energy correlators in jet substructure at the EIC. Using state-of-the-art simulations, we demonstrate that the energy correlators cleanly image QCD dynamics as a function of scale, allowing us to isolate medium modifications from initial-state effects. We then show that the energy correlators are able to achieve femtometer resolution, allowing us to resolve size differences between ${}^3$He, ${}^4$He, ${}^{12}$C, ${}^{40}$Ca, ${}^{64}$Cu, ${}^{197}$Au, and ${}^{238}$U nuclei with EIC kinematics. 

\emph{EIC Simulation and Analysis}---
To illustrate the key features of the energy correlators in electron-nucleus collisions, we employ the eHIJING framework \cite{ehijing}, which is specifically designed to study nuclear-modified jet evolution in DIS. eHIJING combines Pythia 8 \cite{Sjostrand:2014zea,Cabouat:2017rzi} for describing the hard process and the vacuum parton shower, with a description of the nuclear modification by multiple collisions between shower partons and small-$x$ gluons of the nucleus. In eHIJING the transverse momentum dependent gluon density $\phi_g(x, k_\perp^2)$ of the nucleus is parameterized by a saturation-inspired formula~\cite{Dumitru:2011wq}, 
\begin{eqnarray}
\label{eq:gluondensity}
&& \phi_g(x, k_\perp^2) \equiv \frac{K}{\alpha_{s, \rm eff}} \cdot x^\lambda (1-x)^{n} \frac{1}{k_\perp^2+\hat{q}_g L}, 
\label{eq:ehijing:phig}
\\
&& \hat{q}_g = \rho_0 L \frac{C_A}{d_A}  \int_0^{Q^2/x_B} \alpha_{s, \rm eff}~ \phi_g(x, k_\perp^2)  d^2k_\perp,\,\,\,
\label{eq:ehijing:qhatg}
\end{eqnarray}
where $\alpha_{s, \rm eff}$ is an effective jet-medium coupling constant,
$\rho_0=0.17~{\rm fm}^{-3}$ is a reference nucleon density, and $L$ is the path length of the jet's propagation within the nucleus. The gluon jet transport parameter, $\hat{q}_g$, is defined as the average collisional momentum broadening per unit path length of propagation in nuclear matter.  The value of $\hat{q}_g$ is self-consistently determined from the gluon density $\phi_g$, and the average broadening $\hat{q}_g L$ screens the infrared divergence in Eq. \ref{eq:ehijing:phig}. The same nuclear gluon density is then applied to compute the medium-induced QCD splitting function \cite{Wang:2001ifa,Zhang:2021tcc}, which further modifies the evolution of the parton shower and induces radiative transverse momentum broadening.

Most importantly for our study, the parameter $K$ appearing in Eq.~\ref{eq:gluondensity} is an overall normalization factor for $\phi_g$ encoding the magnitude of the medium modification. The choice $K=4, n=4, \lambda=-0.25$ is found to provide a good description of nuclear modification effects in inclusive pion production in HERMES data \cite{ehijing}. In this \emph{Letter}, we will parameterize the magnitude of medium modification directly in terms of the parameter $K$, using a range of values between $2$ to $10$. For reference, this corresponds to a $\hat{q}_g$ in the range of $(0.063, 0.172)$ GeV$^2/$fm at $x_B=0.1, Q^2=1.0$ GeV$^2$ for the average path length of a Au nucleus. The complete dependence of $\hat{q}_g$ on $x_B$ and $Q^2$ for $K$ between $2$ and $10$ can be found in the \emph{Supplemental Material}.  

For our phenomenological study we first generate data for each collision system by producing $4\times 10^8$ events in eHIJING, equivalent to approximately $10$ fb$^{-1}$ of integrated luminosity at the EIC, which should be achievable given the designed luminosity\footnote{An integrated luminosity of $10$ fb$^{-1}$ corresponds to 230 days of running at $10^{33}\text{cm}^{-2}\text{s}^{-1}$,
assuming 50$\%$ running efficiency.} \cite{osti_1765663}. Jet reconstruction is performed using the anti-$k_T$ \cite{Cacciari:2008gp} algorithm with a jet radius of $R$ = 1.0, as implemented in FastJet \cite{Cacciari:2011ma}. All final state particles within $|\eta| <$ 3.5, are used in the reconstruction, and jets clustered from these particles are retained for further analysis. Throughout this {\emph Letter} we use the convention that forward rapidity is along the hadron beam direction, which agrees with the convention used at HERA \cite{Klein:2008di}. This rapidity cut roughly corresponds to the acceptance of the EIC central detector \cite{ABDULKHALEK2022122447, ATHENA:2022hxb, Adkins:2022jfp}. The energy correlator jet substructure observables are constructed using charged jet constituents with $p_T >$ 0.5 GeV.

\emph{Energy Correlators for Cold Nuclear Matter.}---Energy correlator observables \cite{Basham:1979gh,Basham:1978zq,Basham:1978bw,Basham:1977iq,Hofman:2008ar} are statistical correlation functions, $\langle  \mathcal{E}(\vec n_1) \mathcal{E}(\vec n_2) \cdots \mathcal{E}(\vec n_N)  \rangle$, of the asymptotic energy flux. In this \emph{Letter} we focus on the ability of the energy correlators to image nuclear dynamics as a function of scale. It is also worth noting that formulating jet substructure in terms of correlation functions is the long term goal of developing new theoretical techniques to understand extreme QCD matter. See e.g.  \cite{Hofman:2008ar,Belitsky:2013xxa,Belitsky:2013bja,Belitsky:2013ofa,Korchemsky:2015ssa,Belitsky:2014zha,Dixon:2018qgp,Luo:2019nig,Henn:2019gkr,Chen:2019bpb,Dixon:2019uzg,Korchemsky:2019nzm,Chicherin:2020azt,Kravchuk:2018htv,Kologlu:2019bco,Kologlu:2019mfz,Chang:2020qpj,Dixon:2019uzg,Chen:2020uvt,Chen:2020vvp,Chen:2019bpb,Chen:2020adz,Chicherin:2020azt,Chen:2021gdk,Korchemsky:2021okt,Korchemsky:2021htm,Chang:2022ryc,Chen:2022jhb,Chen:2022swd,Lee:2022ige,Yan:2022cye,Yang:2022tgm,Chen:2023wah} for rapid recent theoretical progress in the understanding of the energy correlators.

The simplest correlation functions are one-point functions $\langle \mathcal{E}^n \rangle$, which encode (moments of) the fragmentation spectrum \cite{Caron-Huot:2022eqs}. 
However, one-point functions do not have a scale that can be used to probe the dynamics of jets. One way to overcome this is to consider massive quarks, whose mass naturally introduces a scale \cite{Li:2020zbk,Li:2021gjw}. In nature the quark masses are fixed, so unfortunately this scale cannot be adjusted to probe nuclear scales. 

The scale dependence of physical systems, for example in condensed matter physics or cosmology, is traditionally captured by multi-point correlation functions \cite{Schwinger:1951ex,Schwinger:1951hq}. In the case of asymptotic energy flux, the angular distances between the correlators, $\langle  \mathcal{E}(\vec n_1) \mathcal{E}(\vec n_2) \cdots \mathcal{E}(\vec n_N)  \rangle$, set the scale at which the dynamics is probed. The simplest of these correlators is the two-point function, also referred to as the energy-energy correlator (EEC). The two point function,  $\langle \mathcal{E}^n(\vec{n}_1) \mathcal{E}^n(\vec{n}_2) \rangle$, with $\vec{n}_i$ a unit three-vector specifying the detector direction,  naturally introduces an angular scale $\theta$ from $\vec{n}_1 \cdot \vec{n}_2 = \cos \theta$.\footnote{We use the boost-invariant pseudo-rapidity and azimuthal angle, $\sqrt{\Delta\phi^2+\Delta\eta^2}$, which in the small angle limit is related to the polar angle $\theta$ by the jet pseudo-rapidity, $1/\cosh \eta$. The dependence on pseudo-rapidity cancels out in this limit and thus we use $\theta$ without loss of generality.} Physically this corresponds to the formation time of a splitting, $\tau_f \sim 1/p_T \theta^2$, as illustrated in \Fig{fig:intro}. 

With a weighting power $n>0$, the EEC is ideal for identifying nuclear scales imprinted in the angular structure of collinear radiation within jets. The medium-induced QCD splitting functions of a parton depend on the ratio of the path length of the jet and the formation time of the splitting. 
For splittings where two partons carry comparable energy fraction, this ratio parametrically scales like $L/\tau_f \sim \theta^2 p_T L$, where $L$ is the path length of the jet in the rest frame of the medium. 
Due to the Landau-Pomeranchuk-Migdal interference, splittings with a large formation time $\tau_f\gg L$ (small $\theta$) are strongly suppressed~\cite{Baier:1994bd,Zakharov:1996fv}. Besides, splittings with formation times significantly shorter than $L$ (large $\theta$) are also disfavored because it requires a large transverse momentum transfer induced by pure medium effects, and will eventually be placed outside of the boundary of the jet. Therefore, one expects the onset of medium-induced radiation effects at an angle $\theta_L^2 p_T L\sim 1$, i.e., $\theta_L\propto 1/\sqrt{p_T L} \propto 1/(\sqrt{p_T} A^{1/6})$ with $A$ the mass number of the nucleus. A more refined determination of $\theta_L$ including its dependence on the rapidity of jet and the nucleus is presented in the \emph{Supplemental Material}.
This provides a clean link between nuclear scales and angular scales measured by the energy correlators, as illustrated in \Fig{fig:intro}. This association with the scale of the medium is particularly clean in the case of the EIC as compared with the quark-gluon plasma, due to the fact that the nucleus is a static system, and one can precisely control its size by performing collisions with different nuclear species. In addition to radiative effects, jet-medium collisions can cause a random walk in the parton's transverse momentum, which broadens the angular spread of the radiation pattern with a root mean squared angle $\sqrt{\langle\delta\theta^2\rangle} \sim \sqrt{\hat{q}_g L}/p_T$. For high-$p_T$ jets, one expects $\sqrt{\langle\delta\theta^2\rangle} \ll \theta_L$, and the collisional broadening is subleading to the medium-induced radiative effects. Therefore, we still expect to see the onset of the medium modification to the EEC at $\theta\sim \theta_L$. However, for low-$p_T$ jets and large nuclei, there may be significant broadening effects. 
We leave a more detailed investigation for future studies.

Having understood how nuclear scales imprint themselves into the energy correlators, we now illustrate that this behavior is born out in the eHIJING simulation for realistic EIC kinematics. The expected behavior of the medium modification is clearly visible in \Fig{fig:raw_distros}, where the EEC is shown for several different values of the gluon jet transport parameter $\hat{q}_g$, as parameterized by different values of $K$. We see that the distributions are identical for sufficiently small $\theta$, becoming different at an onset angle set by the radius of the Au nucleus, and with a magnitude proportional to $K$. We will later demonstrate that we can resolve the size of different nuclei. It is worth noting that our observable addresses one of the major challenges of disentangling the initial- and final-state when studying the cold nuclear matter effects~\cite{Chen:2019gqo,Li:2019dre,Li:2020rqj,Ethier:2020way} by being explicitly differential in the correlation angle of the final-state radiation. To confirm this point, we compared the EECs for different nuclei\footnote{Nuclear PDF and isospin effects for different nuclei are included in our analysis.} with $K=0$ and found their shape to be indistinguishable. We believe that this feature of the energy correlators will help in comparing the behavior in cold-nuclear matter and hot nuclear matter (quark gluon plasma) \cite{Andres:2022ovj}, enabling a unified picture across systems. 

\begin{figure}
\includegraphics[scale=0.22]{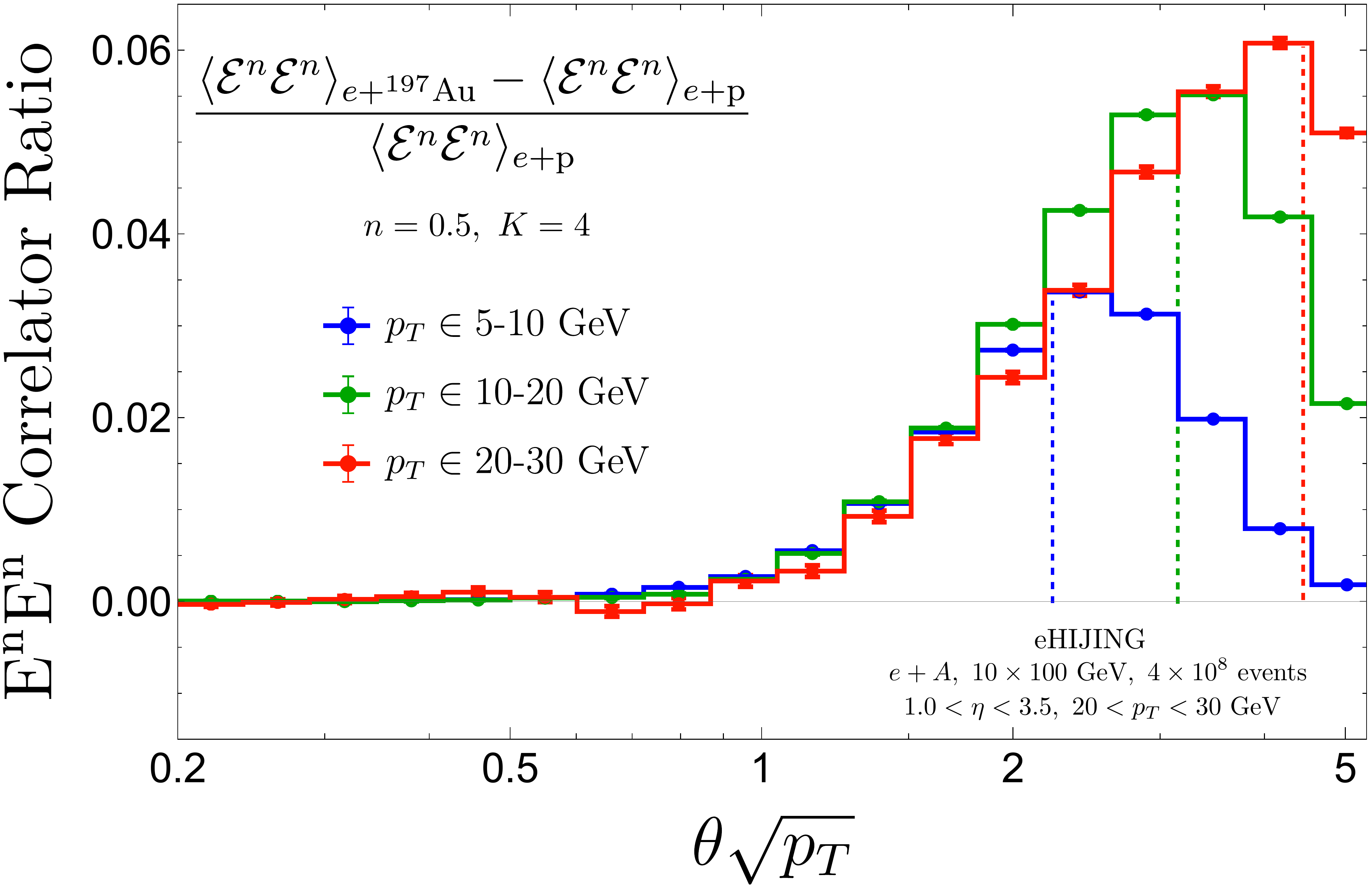}
\caption{Ratios of the EEC on $e$-Au to $e$-proton collisions, for different values of the jet $p_T$,  and $1< \eta<3.5$. The dashed vertical lines denote the jet radius, beyond which the correlator should not be considered. For higher $p_T$ there is a larger region between the initiation of medium effects and the jet radius.
}
\label{fig:pt_scaling}
\end{figure} 

An important feature of the EEC observables is that the onset angle of nuclear modification is set by the size of the nucleus and dimensional analysis, and is therefore model independent. On the other hand, the full shape of the EEC distribution in \Fig{fig:raw_distros} is dependent on the specific features of the eHIJING model of nuclear modification.  Since we focus primarily on the onset angle in this \emph{Letter}, we are confident that our conclusions drawn using the eHIJING simulation are robust. However, we emphasize that there exist various other models to describe jet interactions with cold nuclear matter~\cite{Baier:1994bd,Zakharov:1997uu,Gyulassy:2000er,Guo:2000nz,Guo:2006kz,Vitev:2007ve,Arleo:2003jz,Chang:2014fba,Wang:2009qb,Apolinario:2014csa,Fickinger:2013xwa,Sievert:2018imd,Sievert:2019cwq,Chang:2014fba,Kang:2014xsa,Li:2018xuv,Li:2020rqj,Accardi:2005hk, Kopeliovich:2003py}.   In particular, the first calculation of jet production in the context of electron-nucleus collisions was carried out in~\cite{Li:2020rqj}, where the authors extended the fragmenting jets formalism~\cite{Kang:2016mcy,Kang:2016ehg,Kang:2020xyq} to incorporate initial-state nuclear effects through nuclear PDFs and final-state nuclear effects through medium modified splitting functions. It will be interesting to study the EECs in electron-nucleus collisions using these other approaches in the future.

\emph{Kinematic Dependence.}---We now explore in more detail the kinematic dependence of the EEC. The effect of the medium on the jet evolution depends on the energy $E=Q^2/(2x_B M_p)$, where $M_p$ is the mass of the proton, of the jet in the nuclear rest frame. Here we focus only on the dependence on jet $p_T$ for a fixed $\eta$ bin, leaving more detailed studies to the \emph{Supplemental Material}. All kinematic parameters are chosen to be within the realistic acceptance of the proposed EIC detectors.

In \Fig{fig:raw_distros} we show the normalized distribution of the EEC for electron-proton ($ep$) and electron-Au ($e$Au) collisions. To emphasize the medium modification, it is convenient to take the ratio to the EEC measured in $e$p collisions, which is shown in \Fig{fig:pt_scaling}. This can be done at the EIC, since collisions with different nuclei are measured in a clean and identical collider environment. In \Fig{fig:pt_scaling}, we have chosen to plot the results as a function of $\theta \sqrt{p_T}$ (equivalently $1/\sqrt{\tau_f}$) for three different values of the $p_T$ that can be realistically achieved at the EIC. This reveals that, apart from the overall scaling with $\sqrt{p_T}$, the onset angle is similar for the three $p_T$ bins, with remaining small differences coming from collisional broadening for low-energy jets. After rescaling, the distribution's upper cutoff coming from the jet radius are shifted by $\sqrt{p_T}$. In \Fig{fig:pt_scaling} we have marked the value of this cutoff, computed using the lower bound of each $p_T$ bin,  with dashed vertical lines. Beyond the jet radius cutoff the distribution for the correlator should no longer be studied. We emphasize that this cutoff is not a physical feature of the interaction with the medium, but is entirely due to edge effects from the finite jet radius.

By considering jets with higher $p_T$, we see that we enhance the size of the region between the onset of nuclear modification and the jet radius cutoff as shown in Fig.~\ref{fig:pt_scaling}. Additionally, we observe that the strength of the interaction with the nuclear medium does not depend strongly on $p_T$. Together, this gives rise to a larger overall medium modification for higher jet $p_T$. Higher $p_T$ jets are also preferable theoretically, since they are more perturbative in QCD calculations in the regime of the medium modifications.  Our results suggest that the effects of nuclear modification should be clearly visible for realistic EIC kinematics.

\begin{figure}
\includegraphics[scale=0.20]{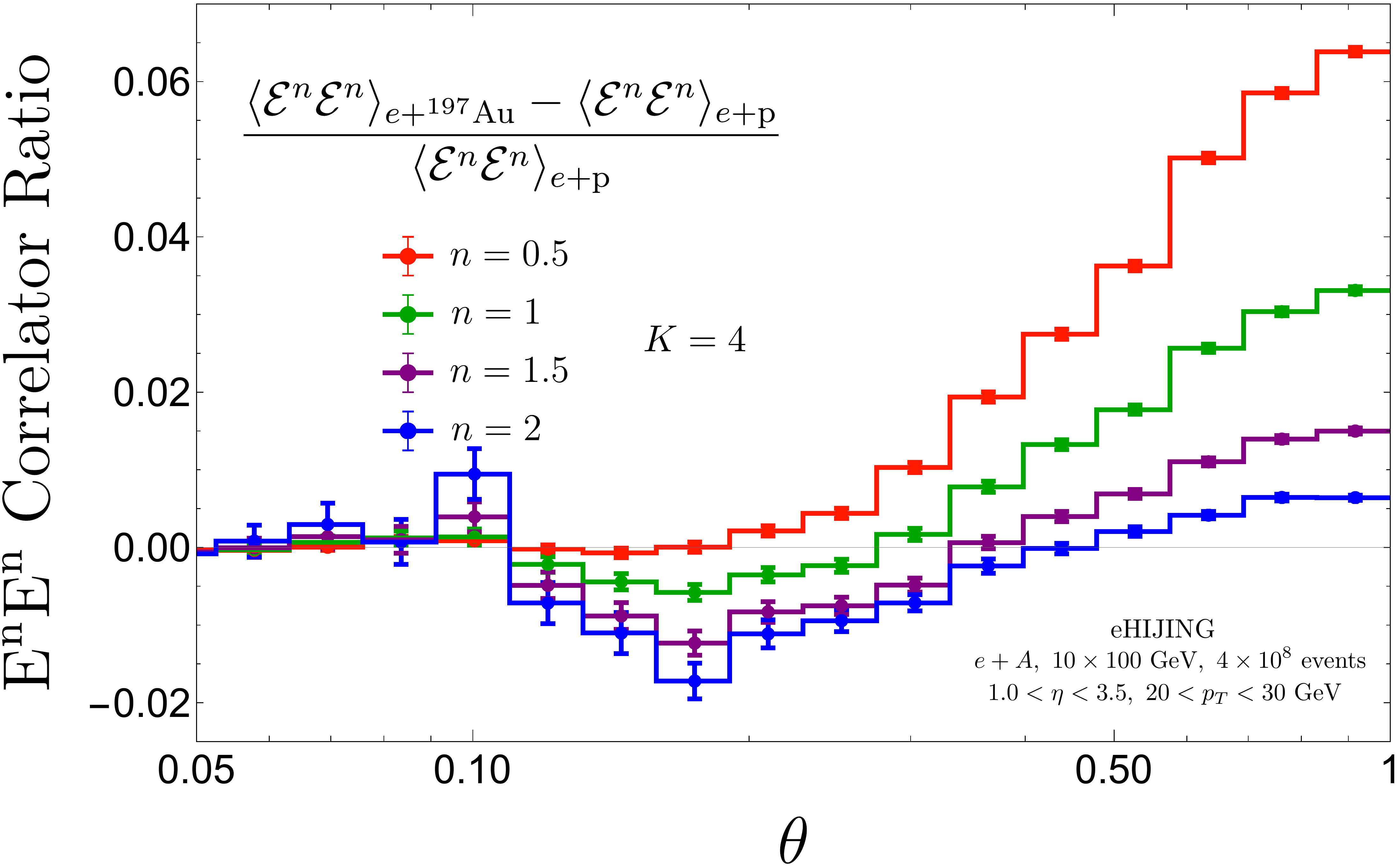}
\caption{Ratios of the EEC, $\langle\mathcal{E}^n \mathcal{E}^n\rangle$, on e-Au to e-proton collisions, for different values of $n$. Smaller values of $n$ enhance contributions from soft radiation, which is more affected by the medium.
}
\label{fig:weights}
\end{figure} 

\emph{Scanning Medium Modification}--- A unique feature of the EIC as compared with hadron colliders is its clean environment. In studies of the energy correlators at hadron colliders \cite{Andres:2022ovj,Holguin:2022epo}, it has been proven useful to consider powers $n>1$ of $\langle \mathcal{E}^n(\vec{n}_1)\mathcal{E}^n(\vec{n}_2) \dots \mathcal{E}^n(\vec{n}_N) \rangle$ to further suppress soft radiation, or backgrounds. For theoretical studies of generalized $E^n$ detectors, see \cite{Caron-Huot:2022eqs}.
Due to the clean EIC environment, it is experimentally feasible to consider the case $n<1$ to enhance soft effects, which are expected to be modified more by the medium. 

In \Fig{fig:weights} we show the medium modified to vacuum ratio of the EEC, $\langle \mathcal{E}^n \mathcal{E}^n\rangle$, for different values of $n$. We observe a much larger medium modification as $n$ is lowered. Studying the dependence on $n$ therefore provides a probe of the energy dependence of medium modification and scans over the energy dependence of the medium-modified QCD splitting functions. It would be interesting to understand this relation analytically.

\emph{Imaging Nuclei.}---A unique feature of the EIC is its ability to provide collisions with a wide variety of nuclei, all with the same experimental setup. Since the EEC probes nuclear scales, we expect to see the sizes of different nuclei imprinted into the correlator. In particular, we expect both the onset angle, as well as the magnitude of the medium modification, to depend on the nucleus species. This clear relationship between the length scale of the nuclear medium and angular scale in the EEC distributions has been illustrated for the case of the quark-gluon plasma in \cite{Andres:2022ovj}, which inspired the current study. It would be interesting to have measurements of the EECs on similar energy jets propagating through both cold and hot nuclear matter, to provide insight into the transport properties of both systems.

\begin{figure}
\includegraphics[scale=0.20]{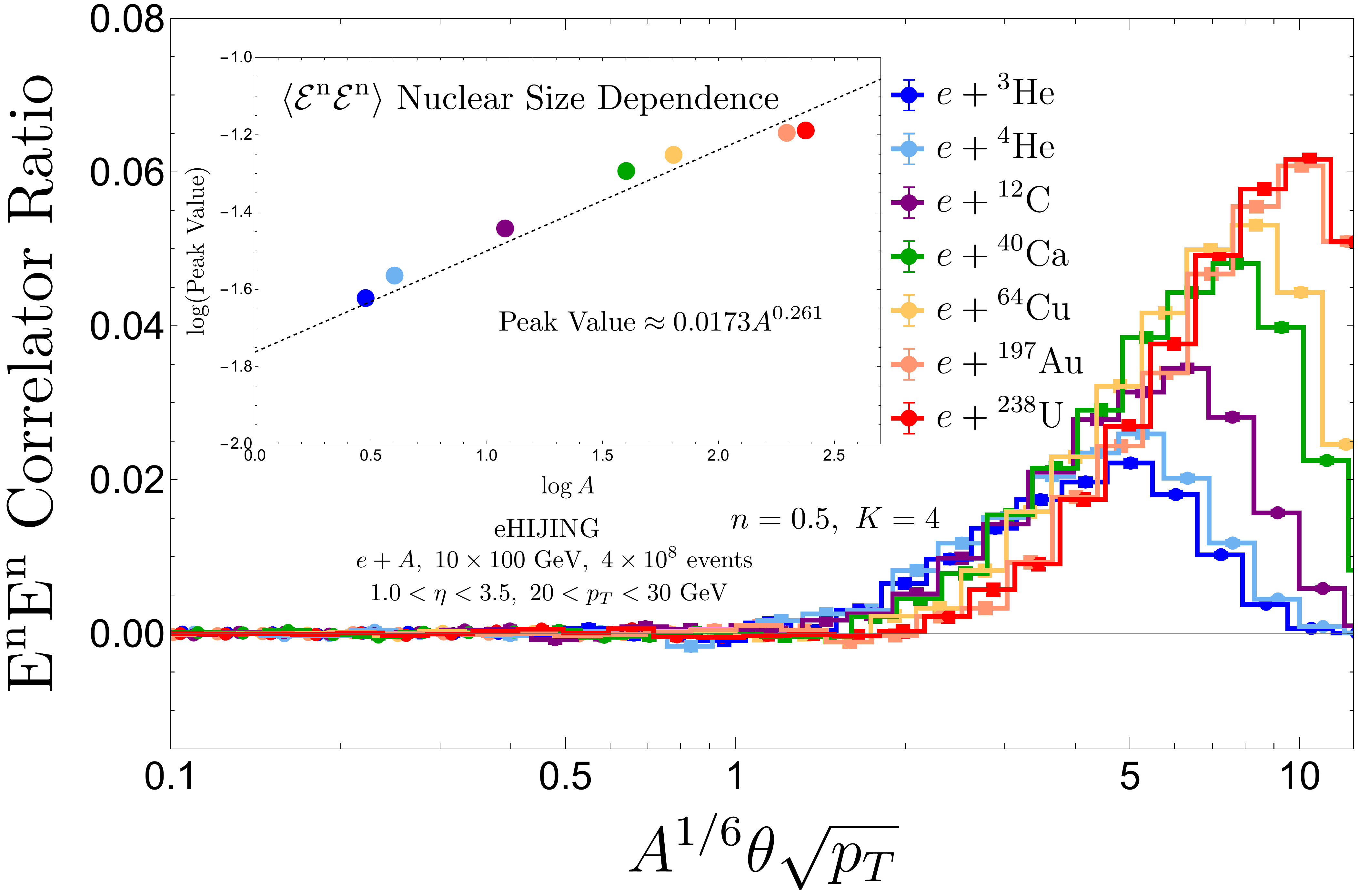}
\caption{The EEC, $\langle\mathcal{E}^n \mathcal{E}^n\rangle$, for various nuclei. The size of the nucleus is directly imprinted into the correlator, as illustrated by the $A^{1/6}$ scaling incorporated into the axis definition. The size of the nucleus can be extracted from the correlator, as shown in the inset. 
}
\label{fig:sizes}
\end{figure}

In \Fig{fig:sizes}, we show the ratio of the EEC to vacuum for ${}^3$He, ${}^4$He, ${}^{12}$C, ${}^{40}$Ca, ${}^{64}$Cu, ${}^{197}$Au, and ${}^{238}$U targets.  By rescaling the x-axis by $A^{1/6}$, the onset of the modification for different nuclei occurs at similar values, which strongly supports the expected scaling behavior of the onset angle $\theta_{L} \propto 1/\sqrt{EL}\propto 1/(\sqrt{p_{T}}A^{1/6})$. The identical plot, but without the rescaling by $A^{1/6}$ is provided in the \emph{Supplemental Material}.  We find it quite remarkable that we are able to achieve femtometer resolution from asymptotic energy flux by using the energy correlators!   We emphasize that the angular scaling of the correlator at which the nuclear modification occurs is set by dimensional analysis, and should therefore be a universal feature of any model of nuclear modification effects. Our analysis shows that this is indeed true for the eHIJING implementation of nuclear modification.  On the other hand, the detailed structure of the EEC in the onset region, and the peak height of the EEC ratio, probe \emph{how} the partons interact with the cold nuclear matter, and are expected to be model dependent. It will be interesting to study the behavior of the EEC observable in different models to provide more insight into the mechanism of nuclear medium interactions. 

One simple observable that we can use as a rough measure of the size of the nuclear modification is the peak height of the EEC ratio. In the inset of \Fig{fig:sizes}, we plot the peak height as a function of $\log(A)$ to study its nucleus size dependence. It shows a beautiful monotonic relation, consistent with a power law scaling.  We expect the details of this relation to be model dependent, and as with the shape of the EEC distribution, it will be interesting to study it in other models of nuclear modification, and ultimately with EIC data.

\emph{Conclusions.}---In this \emph{Letter} we have demonstrated that the energy correlators provide a calibrated probe of the scale dependence of QCD dynamics, and a powerful new way of studying cold nuclear matter effects using jets at the EIC.   Using the eHIJING parton shower, we showed that for EIC kinematics the size of the nucleus is clearly imprinted into the EEC. We also studied the dependence on nucleus size, jet kinematics, and energy weighting of the correlator. Quite remarkably, we were able to observe different nucleus sizes imprinted into the correlator, illustrating a femtometer resolution of the energy correlators.

Building on the extensive recent theoretical and experimental developments in the understanding of energy correlators, there are a number of exciting directions in which this work can be extended. First, it will be important to derive a factorization theorem \cite{Collins:1981ta,Bodwin:1984hc,Collins:1985ue,Collins:1988ig,Collins:1989gx,Collins:2011zzd,Nayak:2005rt} for the EEC in DIS following \cite{Dixon:2019uzg,Lee:2022ige}. This will allow for the calculation of higher order corrections using the known DIS hard functions \cite{Altarelli:1979kv,Furmanski:1981cw,Nason:1993xx,Graudenz:1994dq,deFlorian:1997zj,deFlorian:2012wk,Anderle:2012rq}, as well as a rigorous understanding of the separation of initial and final state effects.  It will also be desirable to compute the energy correlators using theoretical approaches to studying cold nuclear matter effects \cite{Gyulassy:2000er,Gyulassy:2000fs,Gyulassy:2003mc,Vitev:2007ve,Ovanesyan:2011xy,Ovanesyan:2011kn,Fickinger:2013xwa,Ovanesyan:2015dop}, to improve the understanding of how nuclear parameters are imprinted into the correlators. This is of particular interest for comparing calculations of the energy correlators on jets propagating through cold and hot QCD media. The sensitivity of the energy correlators to the angular spread of the radiation in jets suggests that they might be a good probe of collisional broadening effects, and could be useful for extracting $\hat q_g$. 

Another important avenue for future research will be to develop a better understanding of non-perturbative corrections to the energy correlators. While these have been shown to be small for LHC jets, they will be relatively large at the much lower energy scales of the EIC, and their understanding will be necessary for a precision interpretation of measurements. Due to the simple structure of the energy correlator observables, we are optimistic that progress can be made.

Most excitingly, having identified the scale of medium modification using the two-point correlator, one can study higher point correlation functions, providing unprecedented insights into cold nuclear matter. In particular, higher point correlators could allow for a probe of the shape of nuclei and the nuclear medium~\cite{Nijs:2021kvn,Nijs:2022rme,Ebran:2014pda,Busza:2018rrf}.

Combining the proposal of this \emph{Letter} with the recent proposals to use the energy correlators to probe the color glass condensate \cite{Liu:2022wop,Liu:2023aqb}, and the quark-gluon plasma \cite{Andres:2022ovj,Andres:2023xwr}, we look forward to an exciting program using the energy correlators to study QCD in extreme environments in a wide range of collider experiments.

\emph{Acknowledgements.}---We thank Evan Craft, Laura Havener, Barbara Jacak, Ben Nachman, Govert Nijs, Jingjing Pan, Barak Schmookler, Youqi Song, Andrew Tamis, Xin-Nian Wang for useful discussions and comments on the manuscript. We thank Jingjing Pan for help with plots/figures. K.L. was supported by the U.S. DOE under contract number DE-SC0011090. I.M. is supported by start-up funds from Yale University. W.F. is supported by DOE science of office. W.K. is supported by the U.S. Department of Energy, Office of Science, Office of Nuclear Physics through  Contract No. 89233218CNA000001 and by the Laboratory Directed Research and Development Program at LANL.

\bibliography{ref.bib}
\bibliographystyle{apsrev4-1}

\begin{widetext}

\section*{Supplemental Material}
In this \emph{Supplemental Material}, we provide several additional plots further exploring the energy correlators at the EIC, and the jet transport parameters in eHIJING.

\subsection{Nuclei Size Dependence}

In \Fig{fig:nucleisizes}, we show the EEC ratio for different sized nuclei. As compared to \Fig{fig:sizes} where the x-axis was rescaled by $A^{1/6}$, here we have not performed any rescaling. Recall that for the EEC observable, moving to larger angles corresponds to shorter length/time scales. In \Fig{fig:nucleisizes}, we clearly observe that for larger nuclei, the turn on of medium effects starts at smaller angles, and results in a larger total peak height for the medium modification. The alignment of the curves observed in \Fig{fig:sizes} after rescaling by $A^{1/6}$ indicates that these effects exhibit the expected scaling behavior.

\begin{figure}
\includegraphics[scale=0.22]{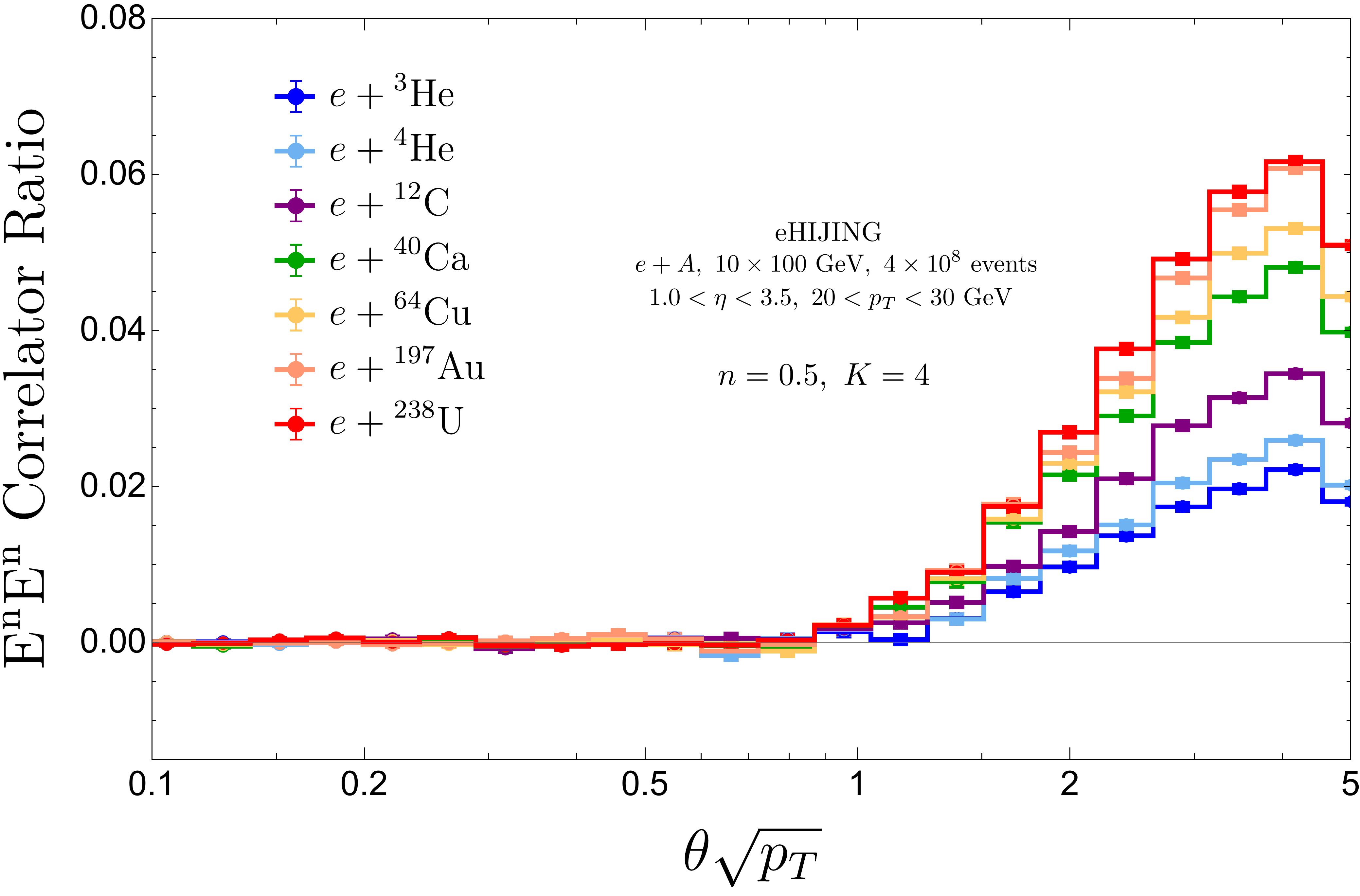}
\caption{The EEC ratio, $\langle\mathcal{E}^n \mathcal{E}^n\rangle$, for various nuclei without the $A^{1/6}$ scaling used in Fig.~\ref{fig:sizes}. As described in the \emph{Letter}, both the onset angle and the peak height depend on the size of the nucleus in the expected manner.
}
\label{fig:nucleisizes}
\end{figure} 

\subsection*{The Jet Transport Parameter in eHIJING}

\begin{figure}
\centering
\includegraphics[width=1.0\textwidth]{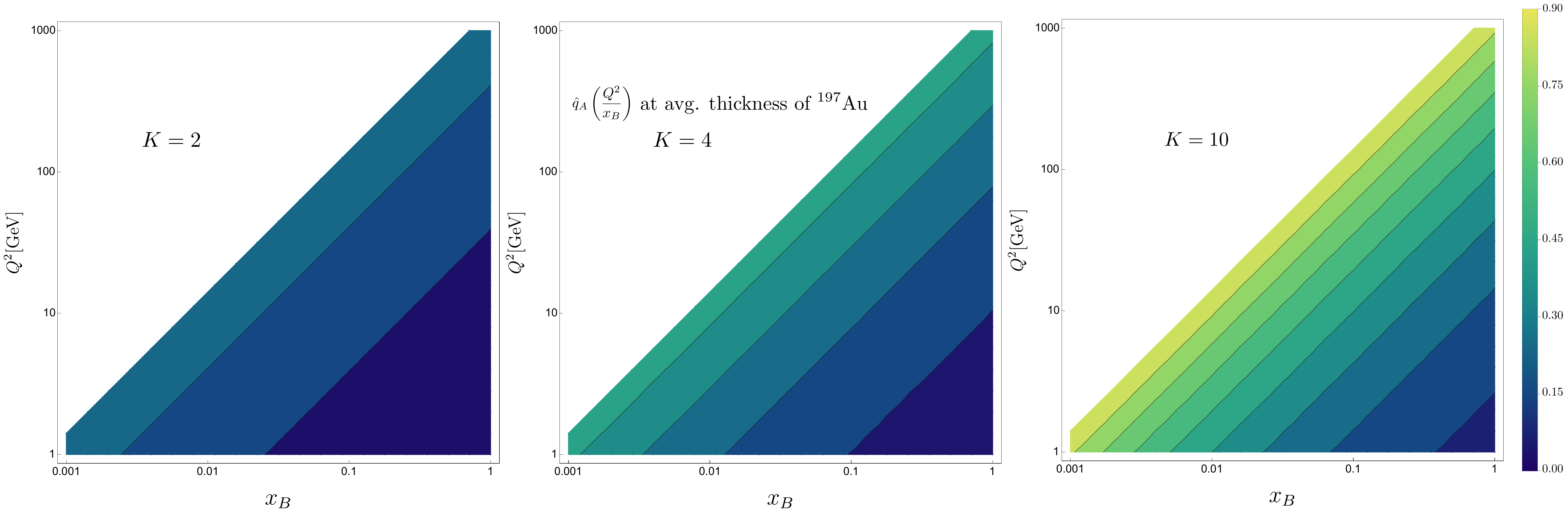}
\caption{The gluon jet transport parameter in units of ${\rm GeV}^2/{\rm fm}$ used in eHIJING as a function of $Q^2$ and $x_B$ with $K=2,4,10$ at the averaged thickness of a gold nucleus. The region outside of the kinematic limit ($Q^2/x_B>s$) is not plotted.}
\label{fig:qhat}
\end{figure}

$\hat{q}_g(x_B,Q^2)$ values for different choices of $K = 2,4, 10$ are shown in Fig. \ref{fig:qhat} at the averaged path length of a gold nucleus. Note that $\hat{q}$ depends on $x$ and $Q^2$ only through the combination $Q^2/(2x_B M_p)$.

\subsection{Characteristic medium angles in the collider frame}
In the small inelasticity limit $(y\ll 1)$, which is the region relevant for this study, the transverse momentum and the pseudorapidity of the jet in the collider can be related to $Q, x_B$ by
\begin{eqnarray}
p_T \approx Q ,~~ \eta \approx \ln\frac{2x_B E_N}{Q} = y_A - \ln\frac{Q}{2x_B M_p}.
\end{eqnarray}
Note that jet moves behind the nucleus in the collider frame.

In the forward rapidity region, where $p_T \ll p_T e^{\eta}\ll m_N e^{y_A}$ and $y\ll 1$. The medium-induced radiation generates a characteristic angle in the EEC distribution, 
\begin{eqnarray}
\label{eq:onset}
\theta_L \sim \frac{e^{\frac{y_A-\eta}{2}}}{\sqrt{p_T L}}
\end{eqnarray}
where $L$ is the path length of the jet in the rest frame of the nucleus. The root-mean-squared collisional broadening of the angle for two splitting partons with energy fraction $x$ (quark) and $1-x$ (gluon) is,
\begin{eqnarray}
\sqrt{\langle \delta \theta^2\rangle} = \frac{\hat{q}_g L}{p_T^2}\left(\frac{C_F/C_A}{x^2}+\frac{1}{(1-x)^2}\right)
\label{eq:rms-angle-coll}
\end{eqnarray}
Again, $\hat{q}_g$ and $L$ are defined in the rest frame of the nucleus. For a nucleon beam energy $E_N=100$ GeV that corresponds to $y_A=4.6$ and a jet with $p_T=20$ GeV produced with $x_B=0.3$, which translates into $\eta\approx 1.1$, $\sqrt{\langle \delta \theta^2\rangle}\approx 0.18$ for symmetric splitting $(x=0.5)$ and $\theta_L\approx 0.27$. These values are comparable to the onset of medium signals in Fig. \ref{fig:raw_distros}.

\subsection{$p_T$ and $\eta$ Dependence}
In \Fig{fig:supp_kinematics}, we present a comprehensive examination of the dependence of the EEC, with weight $n=0.5$, on jet $(p_T, \eta)$. Each panel shows results for different values of the jet transport parameter. In \Fig{fig:heatmap}, we plot the $x_B$ and $Q^2$ distribution of the DIS events for jets in specific $(p_T, \eta)$ bins. As emphasized in the text, the effects from the nuclear medium are largest for large values of $p_T$. In addition, we find that nuclear medium effects are more pronounced for jets produced in forward rapidity region, $\eta\in [1.0,3.5]$, which may seem contradictory as the central and backward rapidity region can access large $\hat{q}_g$ according to the \Fig{fig:qhat}. This is due to the dependence of the Lorentz contraction of the path length on the jet rapidity. In the fixed target frame, the jet path length is given by $L\propto R_A$, where $R_A=1.2 A^{1/3}$~fm is the radius of nucleus $A$. In the lab frame, we obtain rapidity boost factor as seen in the Eq.~\eqref{eq:onset}.
Therefore, as $\eta$ moves backward, the onset angle $\theta_L$ becomes larger, which cannot be accessed with $R\sim\mathcal{O}(1)$. This can be intuitively understood from the fact that the angular scale probed by the correlator in the backward moving jet is required to be larger than the angular scale probed in the forward moving jet in order to ``see'' the forward moving nucleus.  
Meanwhile, from Eq.~\eqref{eq:rms-angle-coll}, the rms angle of collisional broadening evolves slowly with rapidity through $\hat{q}_g(x_B, Q^2)$. Therefore, in the backward region when medium-induced radiation are outside of the probed $\theta$ range, we see pure collisional effects that broadens the EEC signal.

\begin{figure}
\centering
\includegraphics[width=0.93\textwidth]{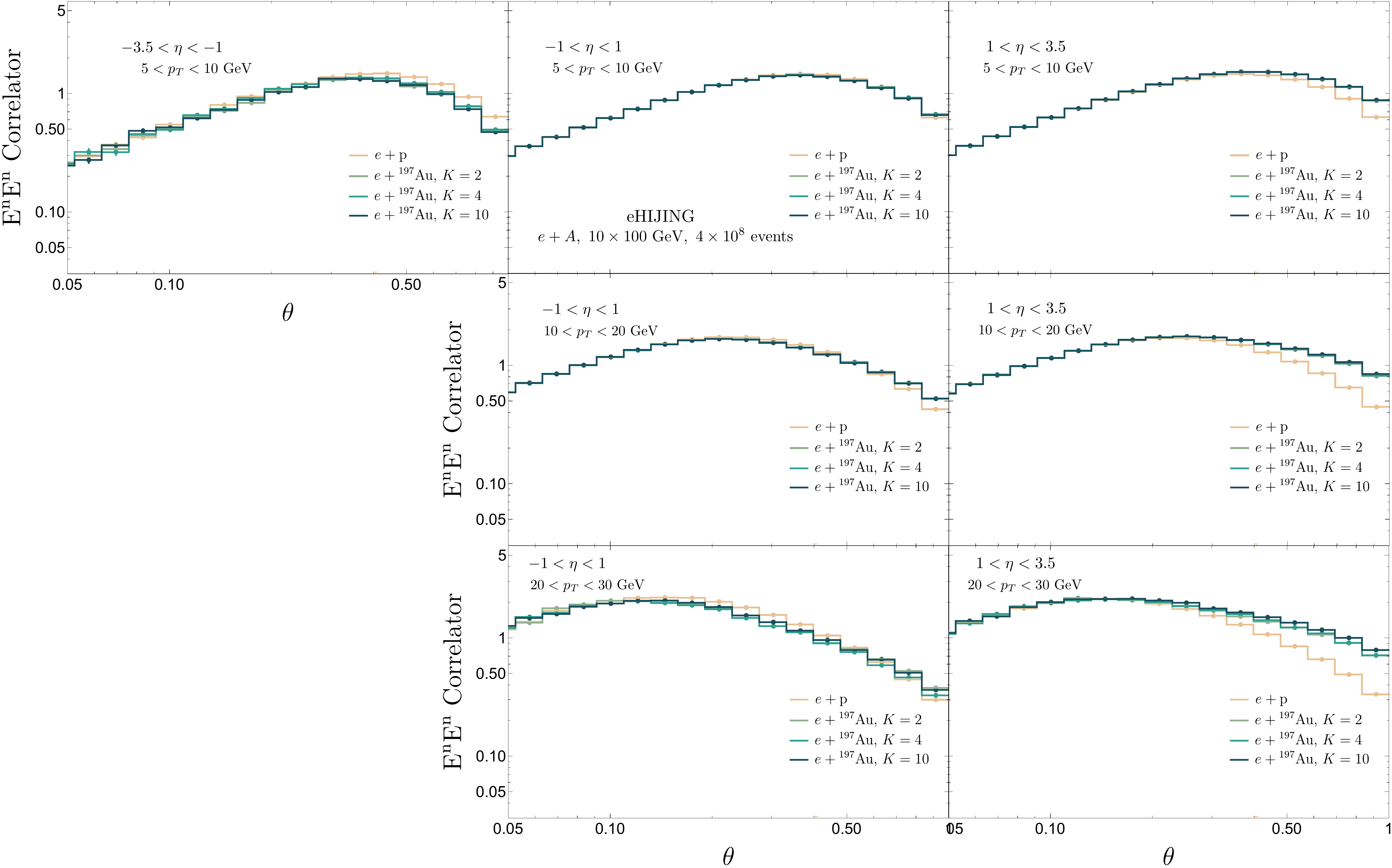} 
\caption{In each panel we plot the EEC, $\langle\mathcal{E}^n \mathcal{E}^n\rangle$, for $n=0.5$, and different values of the jet transport parameter. Different panels illustrate the dependence on the kinematics $(p_T, \eta)$.}
\label{fig:supp_kinematics}
\end{figure}

\begin{figure}
\centering
\includegraphics[width=1\textwidth]{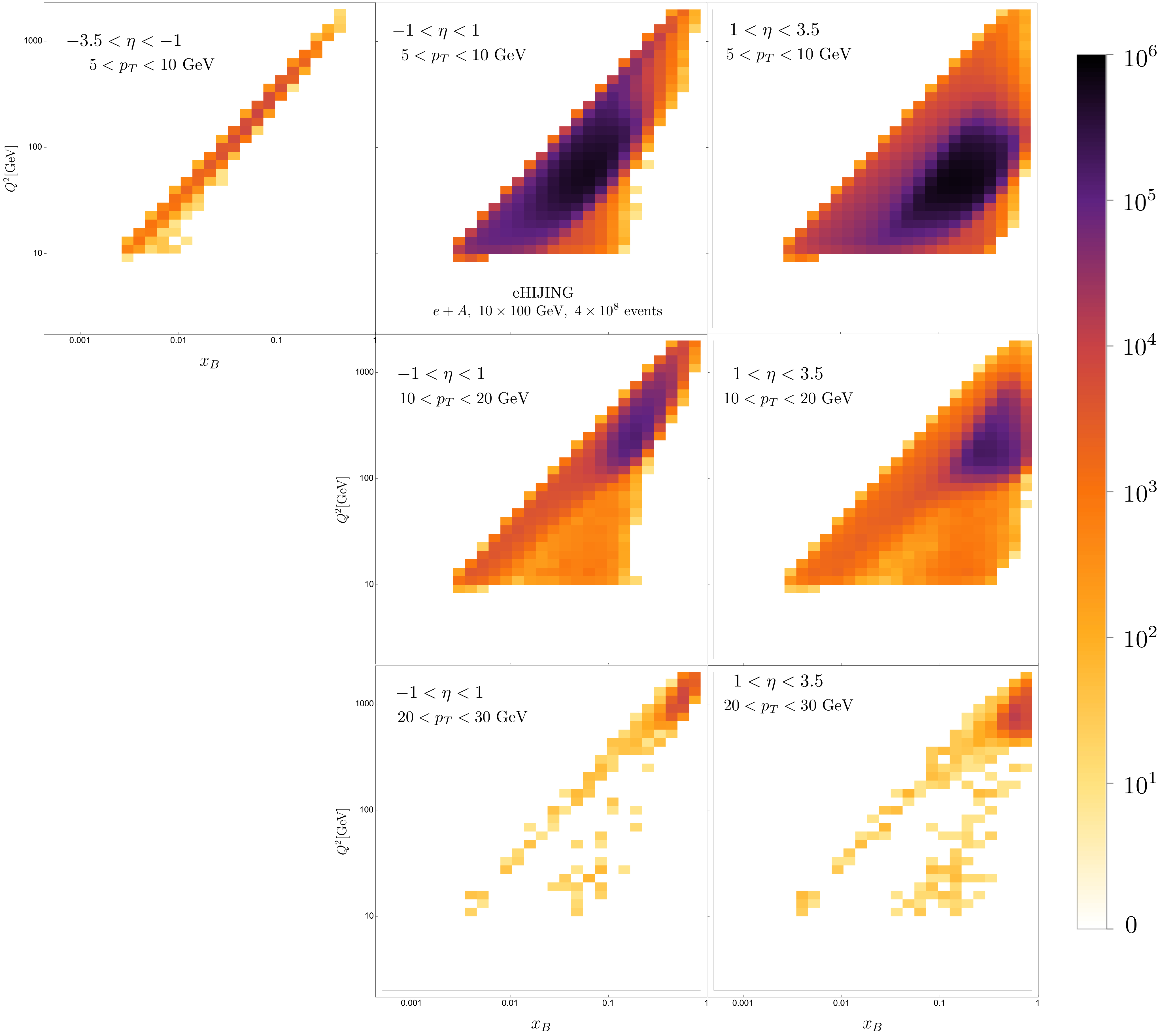} 
\caption{The inclusive jet distribution as a function of  $(x_B,Q^2)$ for the different $(p_T, \eta)$ bins used in Fig.~\ref{fig:supp_kinematics}.}
\label{fig:heatmap}
\end{figure}

\label{sec:supplemental}
\end{widetext}

\end{document}